\documentclass[%
 rsi,
 amsmath,amssymb,
 reprint,%
]{revtex4-1}
\usepackage[dvipsnames]{xcolor}
\usepackage{graphicx}% Include figure files
\usepackage{dcolumn}% Align table columns on decimal point
\usepackage{bm}% bold math
\DeclareUnicodeCharacter{2212}{--}
\DeclareUnicodeCharacter{0308}{\"o}

\usepackage[utf8]{inputenc}
\usepackage[export]{adjustbox}
\usepackage[T1]{fontenc}
\usepackage{mathptmx}
\usepackage{etoolbox}
\usepackage{amsmath,amssymb,amsfonts}
\usepackage[notrig]{physics}

\usepackage{siunitx} %Darstellung von physikalischen Größen
\DeclareSIUnit\year{a} %Neue Einheit
\DeclareSIUnit\val{val}
\DeclareSIUnit\gramm{g}
\DeclareSIUnit\pixel{px}
\DeclareSIUnit\debye{D}
\DeclareSIUnit\sample{S}
\DeclareSIUnit\u{u}

\makeatletter
\def\@email#1#2{%
 \endgroup
 \patchcmd{\titleblock@produce}
  {\frontmatter@RRAPformat}
  {\frontmatter@RRAPformat{\produce@RRAP{*#1\href{mailto:#2}{#2}}}\frontmatter@RRAPformat}
  {}{}
}%
\makeatother
\begin{document}

\preprint{AIP/123-QED}

\title{A scalable scanning transfer cavity laser stabilization scheme\\based on the Red Pitaya STEMlab platform}
% Force line breaks with \\
\author{E. Pultinevicius}
 \affiliation{5. Physikalisches Institut and Center for Integrated Quantum Science (IQST), Pfaffenwaldring 57, Universit\"at Stuttgart, 70569 Stuttgart, Germany}%Lines break automatically or can be forced with \\

 \author{M. Rockenh\"auser}
 \affiliation{5. Physikalisches Institut and Center for Integrated Quantum Science (IQST), Pfaffenwaldring 57, Universit\"at Stuttgart, 70569 Stuttgart, Germany}%Lines break automatically or can be forced with \\

 \author{F. Kogel}
 \affiliation{5. Physikalisches Institut and Center for Integrated Quantum Science (IQST), Pfaffenwaldring 57, Universit\"at Stuttgart, 70569 Stuttgart, Germany}

\author{P. Gro\ss}%
\affiliation{5. Physikalisches Institut and Center for Integrated Quantum Science (IQST), Pfaffenwaldring 57, Universit\"at Stuttgart, 70569 Stuttgart, Germany}

 \author{T. Garg}%
\affiliation{5. Physikalisches Institut and Center for Integrated Quantum Science (IQST), Pfaffenwaldring 57, Universit\"at Stuttgart, 70569 Stuttgart, Germany}

\author{O. E. Prochnow }%
\affiliation{5. Physikalisches Institut and Center for Integrated Quantum Science (IQST), Pfaffenwaldring 57, Universit\"at Stuttgart, 70569 Stuttgart, Germany}

\author{T. Langen}
 \email{t.langen@physik.uni-stuttgart.de}
\affiliation{5. Physikalisches Institut and Center for Integrated Quantum Science (IQST), Pfaffenwaldring 57, Universit\"at Stuttgart, 70569 Stuttgart, Germany}

\date{\today}% It is always \today, today,

\begin{abstract}
Many experiments in atomic and molecular physics require simultaneous frequency stabilization of multiple lasers. We present a stabilization scheme based on a scanning transfer cavity lock that is simple, stable and easily scalable to many lasers at minimal cost. The scheme is based on the \emph{Red Pitaya} \emph{STEMlab} platform, with custom software developed and implemented to achieve up to $100\,$Hz bandwidth. As an example demonstration, we realize simultaneous stabilization of up to four lasers and a reduction of long-term drifts to well below $1\,$MHz per hour. This meets typical requirements, e.g. for experiments on laser cooling of molecules.
\end{abstract}

\maketitle

\section{Introduction}
Laser frequency stabilization is an important component of many scientific investigations and technological applications of atoms and molecules. A large variety of techniques exist to stabilize a laser to an atomic resonance~\cite{Wieman1976,Bjorklund1983,Pearman2002,Corwin98} or optical resonator~\cite{Haensch1980, Drever1983, Black2001,Alnis2008}, ranging from simple spectroscopic schemes to the millihertz-level techniques used to realize the most precise optical clocks~\cite{Zhang2017,Bothwell2022} and gravitational interferometers~\cite{Kwee2012}. 

However, advanced methods for laser stabilization can be complex and expensive, while in many experiments only moderate precision and bandwidth are required. A recent example are experiments aiming to laser cool molecules~\cite{Tarbutt2018, Langen2023,Kogel2021a}. In comparison to atoms, molecules feature a much more complex level structure, where decay into undesired states needs to be compensated through --- sometimes up to almost a dozen~\cite{Vilas2022} --- repumping lasers. At the same time, only moderate compensation against  drifts on the order of the transition linewidths are required, typically in the range of a few MHz. For such experiments, stabilization schemes that are simple, robust and scalable to many lasers are thus highly desirable. 

Here we present a scanning transfer cavity lock which is ideally suited to meet these requirements. Our approach utilizes the readily available, high-resolution analog-to-digital converter and fast digital signal processing capabilities of the \emph{Red Pitaya} (RP) \emph{STEMlab} platform~\cite{RedPitayaWebseite}. Combining this with a simple Fabry-Perot cavity enables long-term sub-MHz-level frequency stabilization using a simple setup, which we demonstrate for up to four lasers in parallel. Our scheme offers a cost-effective and scalable solution for stabilizing lasers that is applicable to a large variety of experiments.

\begin{figure}[tb!]
	\centering
	\includegraphics[scale=0.97]{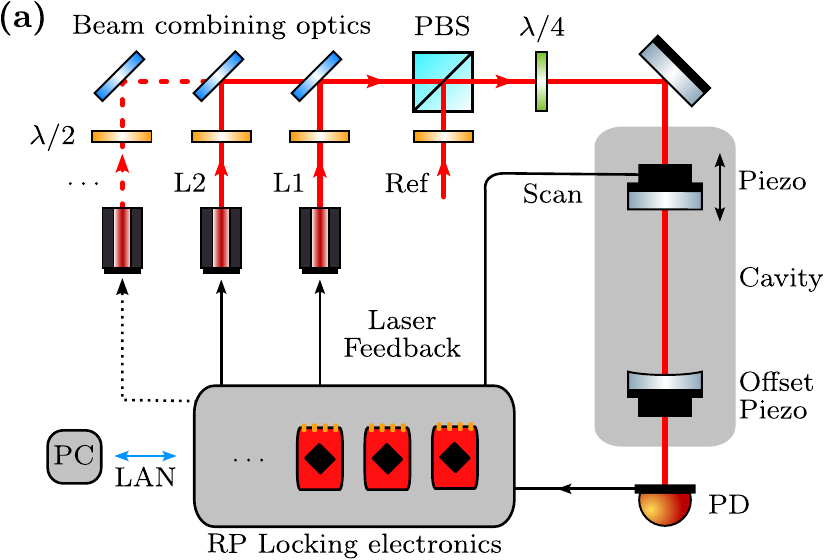}\\
 \vspace{15pt}
    \includegraphics[scale=0.97]{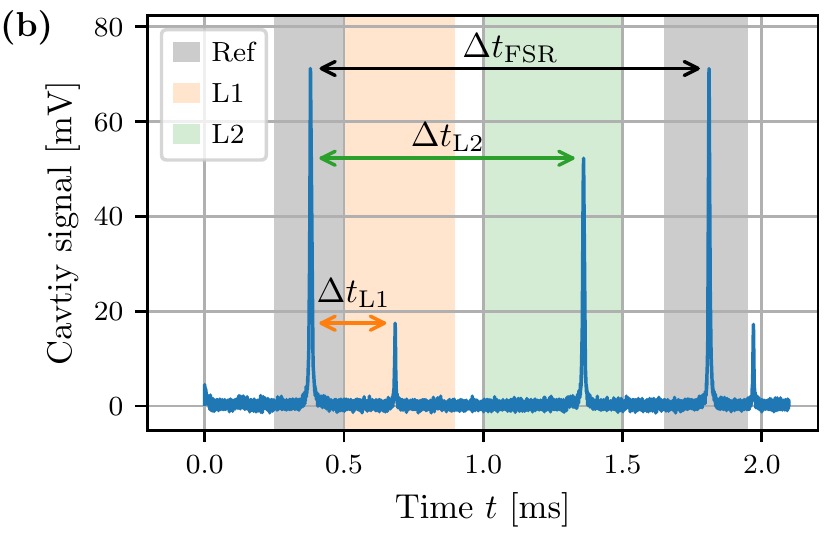}
	\caption{\textbf{(a)} Scheme of the scanning transfer cavity lock. A stable reference laser (Ref) is used to stabilize the length of a scanning Fabry-Perot cavity by recording the laser's cavity transmission signal on a photodiode (PD). Other lasers (L1, L2, or several more indicated by dotted lines) are superimposed with this laser on the cavity through polarizing and dichroic optics, in order to be frequency stabilized. An additional quarter-wave plate minimizes optical feedback from the cavity reflection back into the lasers. \textbf{(b)} Transmission signal of the lasers when scanning the cavity length with the piezo actuator. The peak positions of the reference laser (peaks in gray-shaded regions, separated by one free spectral range of the cavity) provide feedback to stabilize the length of the cavity. This feedback is realized by adding a suitable bias voltage to the scanning cavity piezo to always keep these peaks in the same location. The transmission peaks of any other lasers are detected in their respective regions of interest (color-shaded regions) and their pre-defined, relative distances to the reference laser (indicated by arrows) are kept constant by providing feedback. All feedback signals are generated using synchronized RP units that are remote-controlled by the user through a PC. Before engaging the lock, any potential overlap between the transmission peaks of different lasers can be removed by shifting the wavelength-dependent resonance condition of the cavity by several free-spectral ranges through a second offset piezo actuator (see Appendix A).}
    \label{fig:transfer_cavity_setup}
\end{figure}

\section{Locking Scheme}
The basic setup for our laser frequency stabilization scheme is shown in Fig.~\ref{fig:transfer_cavity_setup}. It is based on the well-established, generic concept of a scanning transfer cavity lock (STCL). Beams from several lasers are superimposed and their transmission through a scanning Fabry-Perot cavity is monitored~\cite{SalehTeichBook}. One of the lasers is frequency stabilized and serves as a reference for the scheme. This initial frequency stabilization can be realized by locking the reference laser to an atomic vapor spectroscopy or by using an inherently frequency-stable laser, such as a helium-neon laser. The positions of the cavity transmission peaks of this reference are then stabilized by changing the cavity length. This transfers the frequency stability of the reference laser to the cavity and stabilizes it against drifts. The positions of all other transmission peaks on this --- now precisely controlled --- frequency scale are subsequently used to generate feedback, by means of which the corresponding lasers are stabilized. 

\begin{figure*}[htb]
	\centering
		\includegraphics[scale = 1]{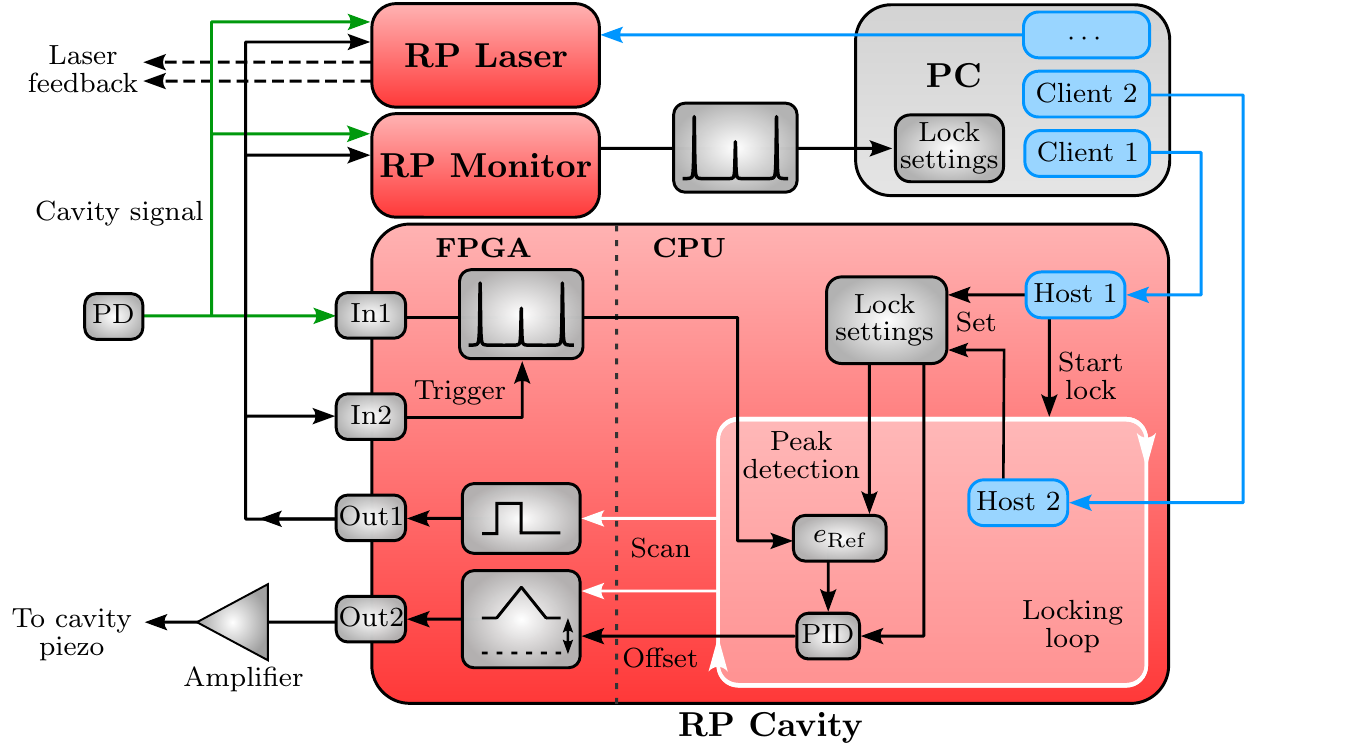}	
	\caption{Schematic of the signal acquisition and processing of the STCL utilizing the RPs and remote communication. The cavity scan is performed by RP Cavity. It is started remotely using a socket connection on Host 1. This starts a loop which blocks said communication. For that reason, an additional socket connection (Host 2) is used to remotely control the running locking loop. The signal generation and acquisition is carried out by the FPGA, while the signal processing for feedback generation happens digitally on the CPU (for details see main text). The configuration depicted is suitable to stabilize two lasers using the a single RP unit (RP Laser). It can be expanded to stabilize up to $2N$ more lasers using $N$ additional RP Laser units. A third RP unit (RP Monitor) or an oscilloscope can be used to monitor the cavity signal.}
	\label{fig:STCL_Remote_sketch}
\end{figure*}

\section{Implementation}

Similar STCL schemes have been widely used in many experiments in atomic and molecular physics~\cite{Burke2005,Seymour-Smith2010,Subhankar2019,Wang2018,matsuhimaPhD,BarryPhD}. Here, we employ RP units for a simple, cost-effective and fully digital processing of the transmission data of the transfer cavity. 

The RP is a multi-instrument, FPGA based platform with open-source capabilities. Fast analog-in and output channels make it very suitable for the processing of analog signals. In the context of laser stabilization, a powerful toolkit is provided for this platform by the \emph{PyRPL} \emph{python} package~\cite{Neuhaus2017}. In terms of cavity stabilization, this package provides, e.g., an interface for analog-to-digital conversion, signal processing and generation. In principle, this makes the package suitable for realizing a STCL, but it also requires permanent data transfer between the RP platform and a personal computer~(PC) for signal processing that is not carried out on the FPGA itself. If applied to setups with many lasers, this typically creates a large data transfer overhead that limits the realizable bandwidth of the stabilization. Our solution aims to minimize this data overhead, by performing all processing and monitoring tasks directly on the RP's internal CPU. It also minimizes the complexity of the optical setup by using only a single cavity for a potentially large number of lasers, and does not require any additional radio-frequency components.

To realize the locking, several RP units are operated in a stand-alone way, where only the basic control commands to run and control the STCL locking loop are externally provided by the user via a local area network (LAN) connection. This connection is realized using \emph{Python} modules that are run on both the RP units and the PC. In our scheme, we use a script to host servers on the RP via \emph{Python's} \emph{socket} package. We then employ an interface available on the RP for the use in \emph{Jupyter} notebooks, which provides access to the FPGA functions. With that, the analog input and output channels can be controlled directly through the RP platform's CPU. 

A schematic illustration of components operated on the RP units, on the controlling PC, as well as their mutual communication is shown in Fig.~\ref{fig:STCL_Remote_sketch}. The basic setup requires one RP (\emph{RP Cavity}) for the scan and stabilization of the transfer cavity to the reference laser. Every pair of additional lasers on the same cavity then requires one additional device (\emph{RP Laser}), since each RP features two fast analog outputs and can thus provide feedback for two lasers. 

The cavity is scanned by bursts of triangular signals, generated using the signal generation capabilities of RP Cavity's FPGA. The frequency of the waveform (\SI{238}{\hertz}) is chosen such that the period is a multiple of the minimum acquisition duration of the FPGA's inputs (In1 and In2). This way the acquisition time can be matched to the duration of the upwards slope. During the signal processing, the cavity piezo can settle back to its original length throughout the downwards slope. The signal acquisition and processing is triggered on all devices using a square wave, synchronized to the cavity scan and generated on the second output of RP Cavity.

The detected cavity transmission is evaluated on the CPU of RP Cavity. This involves the determination of different resonance positions that need to be distinguished from one another. For that purpose, regions of interest are specified for the respective lasers in the cavity (see Fig.~\ref{fig:transfer_cavity_setup}b). The peaks are then detected by searching for the maximum positions in the respective ranges. Using the data points next to those positions, a Savitzky-Golay derivative filter~\cite{Schafer2011} is applied. The resulting zero-crossing is determined using linear regression to find the resonance position~\cite{Subhankar2019}. This method efficiently smoothens the data around the resonance and increases the time resolution of the peak detection algorithm. The ranges of interest as well as the necessary parameters for the peak-detection algorithm are remotely provided from a PC and are saved on the device.

From the evaluated peak positions for the reference laser and the lasers $\text{L}_i$ the relative positions $\Delta t_{\text{L}_i}$ (see Fig.~\ref{fig:transfer_cavity_setup}b) are calculated to obtain the respective error signals

\begin{equation}\label{eq:Error}
    e_{\text{L}_i} = \frac{\Delta t_{\text{L}_i}}{\Delta t_{\text{FSR}}} - s_{\text{L}_i}.
\end{equation}

The set values $s_{\text{L}i}$ are defined by the user. Variations of the scan ramp are cancelled out by the division over the FSR $\Delta t_{\text{FSR}}$ of the cavity~\cite{Subhankar2019}. This includes effects that can arise due to the non-linearity and hysteresis of the cavity piezo. For cavity stabilization, the error $e_\text{Ref}$ is simply defined as the distance of the resonance to a pre-defined set value.

In addition to the signal processing, the locking also requires a feedback loop, which is not directly available on the RP's FPGA. We thus implemented a dedicated digital proportional–integral–derivative controller (PID controller). As the aim of our locking scheme is to compensate against slow long-term drifts of laser frequencies, we use mainly the integral (I) and proportional (P) terms of this controller.

A drawback of our remote-controlled implementation of the lock is that the actual data on the RP can not be accessed while the lock is enabled without compromising locking bandwidth. For this reason an additional RP unit (\emph{RP Monitor}) that is independent of the lock is used for monitoring of the cavity signal on the computer. This unit runs the same algorithms as the units used in the transfer lock, which assures that the monitored signal is as close as possible to the actual STCL signals. To prevent the transfer of relatively large data sets between RP Monitor and the user's PC, the errors of the individual peak positions can be evaluated on the unit before they are sent to the computer. This allows for faster update rates when monitoring the locking performance. As an alternative to this additional unit, the cavity signal can also simply be monitored with any oscilloscope.

Overall this leads to a full setup of $2 + \mathrm{ceil}(N/2)$ devices for $N$ lasers on a single cavity, including RP Monitor. On each of the RPs, a continuously listening server is started by remotely running a script. Communication with these servers can be established from a PC, such that commands can be sent to the devices to control the cavity scan as well as the individual locks. Based on the monitored cavity signal, the respective setpoints and peak detection ranges of the individual locks can be initialized. The locks of the RPs can be individually started, which initiates a feedback loop on each device. Those loops block the communication on the original listening server. In order to keep the locks remotely controlled, a second host server is started on the RPs to listen for further commands. This is realized using multi-threading and multi-processing, such that it is possible to adjust any lock settings while the STCL is running. This approach also allows for slow frequency scans within the pre-defined peak detection ranges.

An important benchmark for our scheme is the achievable bandwidth. Taking into account the computing power of the RP and the efficiency of our implementation, we generate feedback output up to every \SI{10}{\milli\second}, i.e. with a bandwidth of up to \SI{100}{\hertz}. This is a comparable timescale to the scan rates used for the cavity. For comparison, it is also an order of magnitude faster than an implementation we realized using the readily available components of the \emph{PyRPL} package. We observe that the main contribution to the time budget of each iteration originates from the acquisition and processing of the cavity signal. In particular, a time delay of around \SI{4}{\milli\second}, independent of the read-out buffer size of the RP, can be attributed to the conversion of the memory buffer to a \emph{Python} object.

A possible improvement of the bandwidth could thus be realized by implementing the whole feedback loop in \emph{C} or directly on the FPGA. We estimate that this could increase the bandwidth by another order of magnitude, up to the limit of typical piezo actuators, which is in the kHz range~\cite{Seymour-Smith2010}. However, since our main goal in this work is to minimize slow, long-term drifts of the lasers, we favor the use of \emph{Python}, which results in more accessible code. 

\section{Operation of the lock}

\subsection{Optical setup}\label{sec:opticalsetup}
We tested the STCL using several realistic setups for the laser cooling of atoms and molecules. This involved different home-built and commercially available external cavity diode lasers of different wavelengths, as well as a tunable continuous-wave Ti:Sapphire laser. To address these different lasers, feedback of different forms was generated using the RP units. For the diode lasers, the feedback was applied through their diode currents or through piezo actuators controlling the length of an external Littrow cavity. In cases where only small feedback is required and which may thus be limited by the $14\,$bit resolution of the RP's output signal, this signal can be improved by combining higher output amplitudes with an attenuator. For the Ti:Sapphire laser the feedback was applied to a piezo actuator, which controls the length of an internal cavity of the laser. 

For the transfer cavity, we typically used a simple home-built design in our tests~\cite{Tomschitz2018}. It consists of a plane and a confocal mirror, which are separated by \SI{165.4}{\milli\meter} using a stainless steel spacer, resulting in a free spectral range of \SI{906}{\mega\hertz}. Given a particular wavelength range of interest, we used custom mirror coatings, resulting in a typical finesse on the order of $240$. The mirrors of the cavity are mounted on piezo ring actuators, allowing for fine control of the resonator length. As indicated in Fig.~\ref{fig:transfer_cavity_setup}a, the first piezo actuator was used for scanning and stabilization of the cavity length, while the second was used to realize a fixed length offset independently of the operation of the STCL. 

This latter feature is useful to easily minimize the overlap of transmission peaks from different lasers and is thus crucial for the scalability of our scheme. As the transmission resonance condition for each laser depends on its wavelength, any incidental overlap between two peaks from different lasers can be removed simply by shifting the cavity length by several free-spectral ranges (see Appendix~\ref{appendix:shiftingpeaks}). The same can also be achieved with a single piezo in combination with a bias voltage on the scan ramp.

It should be noted that the analog outputs of the RP are limited to a range of $\pm \SI{1}{\volt}$, which is not sufficient to drive a typical piezo actuator over its full range. For this reason, whenever necessary, we employed amplifiers to scale the voltages to sufficient amplitudes~\cite{Mausezahl2023}. In the case of the transfer cavity, this solution not only allows for a variable gain and offset, which is useful for the initialization of the scanning range, but also reduces noise on the scanning ramp. Additional noise reduction is possible using a low-pass filter that is matched to the capacity of the piezo and the speed of the scan ramp. 

Our scheme is not specific to a particular cavity design and can easily be realized also with any other cavity. In particular, cavities mounted in vacuum or constructed using low thermal expansion materials could further improve the performance of the lock presented in this work. 

\begin{figure}[tb!]
	%\centering
	\includegraphics[scale=.95]{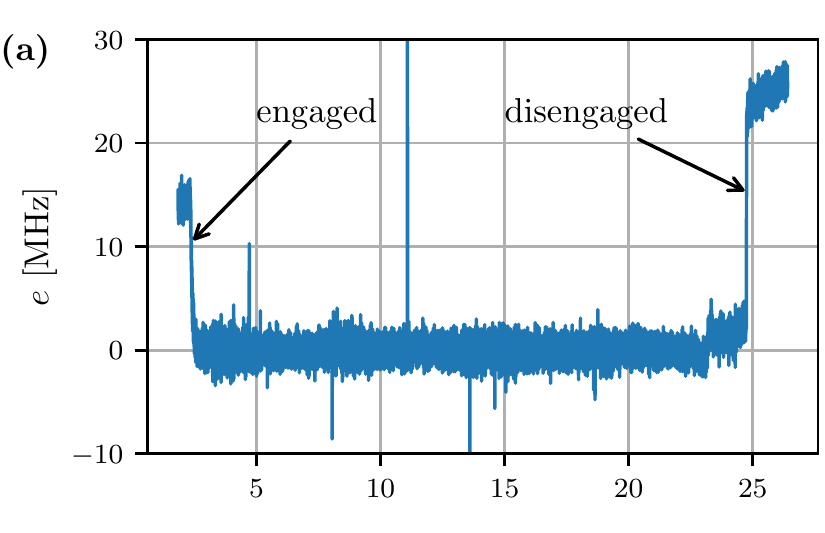}
 \includegraphics[scale=.95]{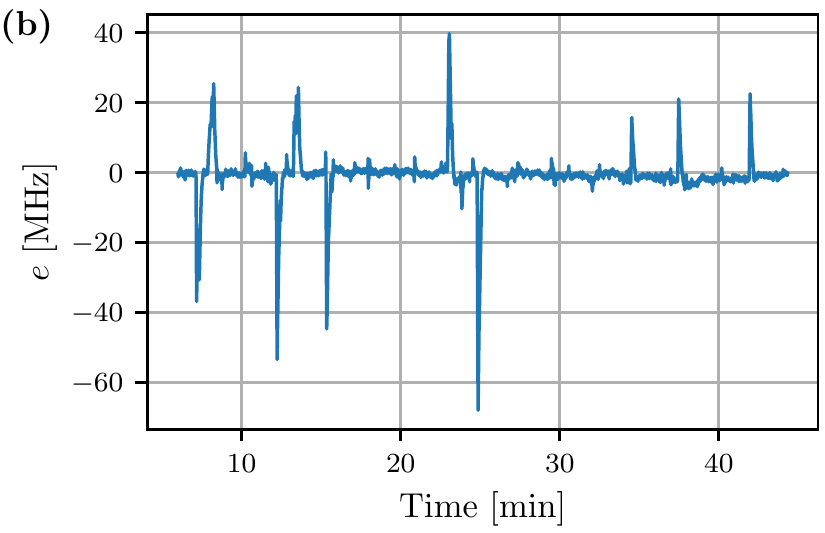}
 \includegraphics[scale=.95]{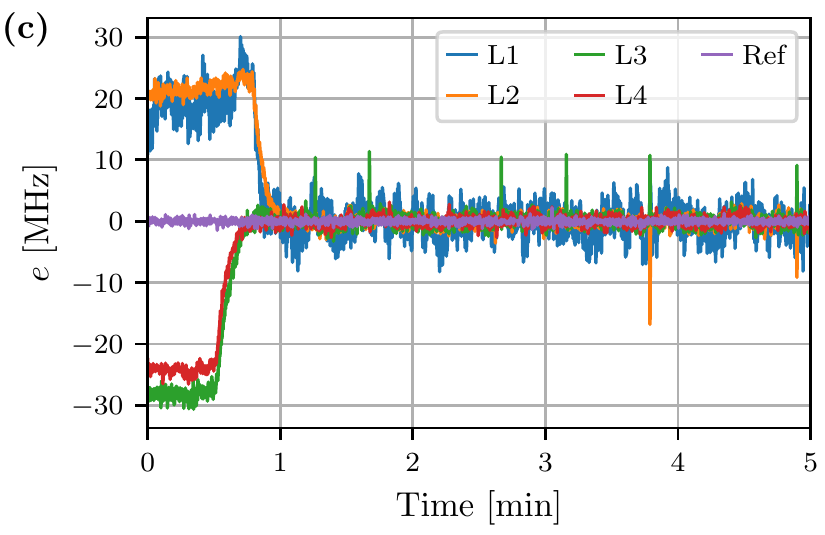}
 \caption{\textbf{(a,b)} Example locking of a single laser. \textbf{(a)} Deviations from the setpoint measured on the transfer cavity over time. After the lock is engaged, the initial error rapidly decreases During the lock, the error fluctuates around zero with a standard deviation of \SI{0.91}{\mega\hertz}. After more than \SI{20}{\minute} the lock is disengaged. Spikes are the result of small external perturbations which are rapidly compensated by the STCL. \textbf{(b)} Performance of the lock while the laser is purposely subjected to major external perturbations. Again, after each perturbation the lock rapidly returns the laser to the set frequency. This data was recorded using a wavemeter, which confirms the error evaluation performed on the RPs.  \textbf{(c)} Operation of the STCL for four lasers. The fast noise visible at L1 is the result of a particularly noisy homemade diode laser, which, like L2-L4, could be successfully stabilized.} 
	\label{fig:locktest1}
\end{figure}

\begin{figure*}[tb]
    \centering
    \includegraphics[scale=0.95]{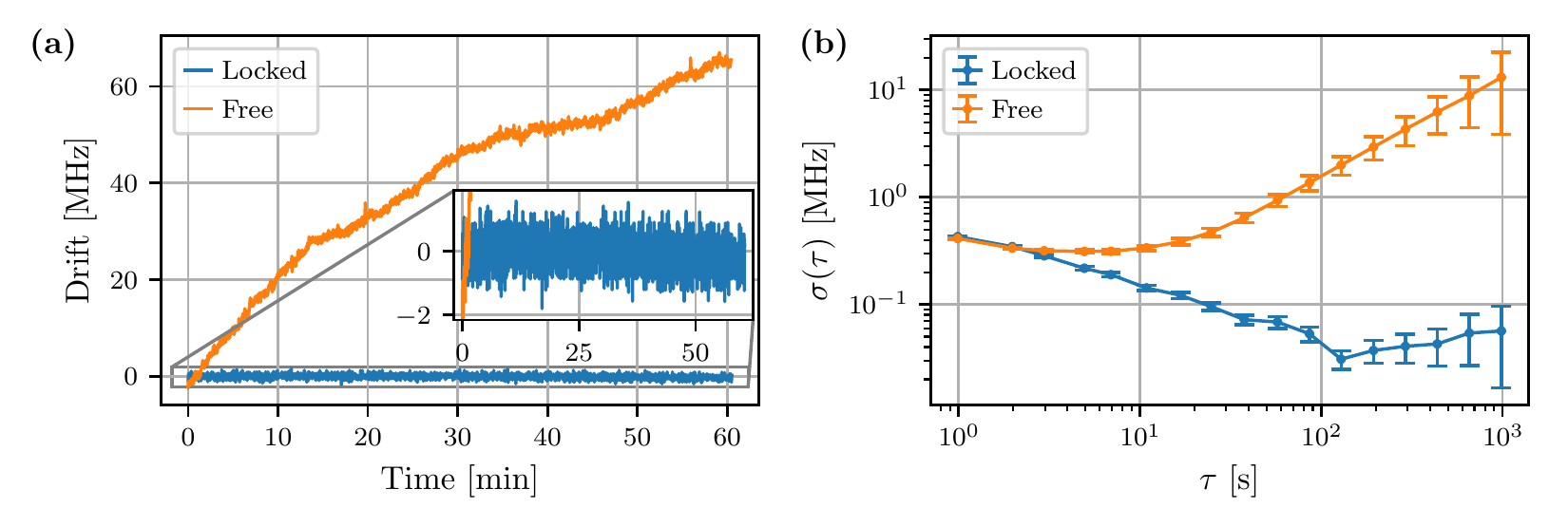}
    \caption{Long-term frequency stability of a laser that is locked by the STCL. \textbf{(a)} Frequency drifts of a laser with (Locked) and without (Free) the STCL engaged over the course of an hour. This data has been evaluated from beat-note measurements relative to an additional frequency-stabilized reference laser. Significant frequency variations of the laser are suppressed by the locking system. \textbf{(b)} Allan deviation $\sigma (\tau)$ of the frequency stabilized laser in (a) for increasing averaging times $\tau$.}
    \label{fig:beat_note}
\end{figure*}

\subsection{Basic locking}\label{sec:basiclock}

An example result for the stabilization of a home-built diode laser is illustrated in Fig.~\ref{fig:locktest1}a. This laser was set close to a transition required for the laser cooling of barium monofluoride molecules that is located at \SI{898}{\nano\meter}~\cite{Albrecht2020}. For the reference, a commercial diode laser was stabilized to the cesium D1 line at \SI{895}{\nano\meter} via a vapor cell FM spectroscopy~\cite{Bjorklund1983}, which was implemented using \emph{PyRPL}~\cite{Neuhaus2017}. 

The error $e$ of the laser frequency was evaluated from the cavity signals as in Eqn.~\ref{eq:Error}, with the frequency scale obtained by multiplication with the free spectral range. In addition to the internal monitoring of the STCL, the frequency scale and the operation of the lock were also independently observed and validated using a wavelength meter throughout the tests. 

In order to illustrate the difference between the locked and unlocked states, the lock was intentionally operated for a time of around $20\,$ minutes.  Frequency jumps on the order of \SI{15}{\mega\hertz} indicate the times when the lock was manually engaged and disengaged. Residual high frequency noise visible on the signal was due to the inherent noise of the home-built laser. We find that the maximum time for which the lock can operate is mainly limited by the thermal drifts of the cavity, which can at some point exceed the tuning range of the piezo actuators. In a temperature stabilized labspace, we have observed this time to exceed several hours. 

\subsection{Stability against perturbations}\label{sec:perturbations}

We further investigated the effects of sudden perturbations on the performance of the STCL using a similar setup as in the previous section. For this, the locked laser was strongly perturbed by manually knocking the optical table or by adding current spikes to the diode laser current while the lock was engaged. The resulting monitoring signal is shown in Fig.~\ref{fig:locktest1}b. Occasional jumps in the signal reach \SI{70}{\mega\hertz} and correspond to the times when the laser was perturbed. Such jumps were well within the pre-defined region of interest for the peak detection of this laser. Therefore, the STCL could always detect the resonance, the resulting frequency spikes were quickly reduced again and the laser remained locked throughout the test, without any significant drift. 

We note that any perturbations larger than the region of interest, such as laser mode-hops, can still interrupt the lock. This is particularly important to keep in mind when more lasers are added to the same cavity, which limits the size of the region of interest available for each laser.

\subsection{Scalability}\label{sec:scalability}

In order to test the scalability of our STCL scheme, we locked up to four lasers simultaneously on a single transfer cavity. Including the monitoring RP, a total of four RP units were used for the frequency stabilization of the lasers L1 to L4. For demonstration purposes, the resonances of the lasers were spaced equally throughout the scanned range. The wavelengths corresponded again to the cooling scheme for barium monofluoride~\cite{Albrecht2020}, which features transitions between \SI{860}{\nano\meter} and \SI{898}{\nano\meter}. The monitored error signals are summarized in Fig.~\ref{fig:locktest1}c, with only the first $5$ minutes shown for clarity. All four lasers were successfully stabilized against any relevant frequency drifts on timescales exceeding hours.

\section{Performance}\label{sec:long_term}

To characterize the long-term performance of the STCL, we used a beat note of two overlapped lasers. An example is given in Fig.~\ref{fig:beat_note}, where the two lasers were realized using commercial diode lasers. One of these lasers, frequency stabilized via DAVLL~\cite{Corwin98} to the D2-line of ${}^{85}\text{Rb}$, served as a precise reference for residual drifts of the second laser. This second laser was locked using the STCL to a frequency roughly \SI{60}{\mega\hertz} away, allowing for a measurement of the beat-note signal using an oscilloscope with \SI{200}{\mega\hertz} bandwidth. For the reference laser used in the STCL scheme, a separate third laser at the same wavelength was used, stabilized with a DAVLL scheme on the ${}^{87}\text{Rb}$ D2 line. The use of two lasers with wavelengths close to each other reduced drifts caused by environmental factors for the STCL~\cite{Subhankar2019}. Moreover, the choice of a separate reference laser for the beat-note measurement assured that the measurements were not influenced by drifts of the reference laser.

The beat-note signal was acquired with a frequency of \SI{1}{\hertz} over the course of an hour. Variations in the frequency of the laser were evaluated by the peak positions of the Fourier-transformed signals. The resulting frequency drift $\Delta f$ is shown in Fig.~\ref{fig:beat_note}a. Without engaging the transfer lock, the frequency of the laser drifts by more than \SI{60}{\mega\hertz} over the course of one hour. These drifts were fully compensated while the STCL was engaged. Notably, the frequency drifts over an hour were characterized by a standard deviation of \SI{0.5}{\mega\hertz}, well below the short-term fluctuations of the individual lasers. The largest residual drifts observed in a large set of such test measurements performed over several weeks of operation were on the order of \SI{1.5}{\mega\hertz}. The Allan deviation calculated from the measured frequency drifts $\sigma(\tau)$ is shown in Fig.~\ref{fig:beat_note}b. For increasing averaging times $\tau$ up to \SI{200}{\second}, $\sigma(\tau)$ decreases due to averaging of short-term noise. For longer $\tau$, an upwards trend indicates the effect of the small residual frequency drifts.

Substituting the STCL's reference with a frequency stabilized helium-neon laser resulted in a similar performance with drifts of \SI{0.8}{\mega\hertz} over three hours, which can be explained by both a temperature drift around \SI{0.5}{\degreeCelsius} during the measurement and the specified frequency stability of the helium-neon laser of \SI{1}{\mega\hertz} per hour.

\section{Conclusion}
We have used the capabilities of the \emph{STEMlab} platform provided by \emph{Red Pitaya} to develop a STCL that is both cost-efficient and scalable. To prevent an overhead in data-transfer, \emph{Python} modules were developed to implement the signal processing and feedback loop digitally directly on the RP's CPU with a bandwidth of up to \SI{100}{\hertz}. The STCL is remotely accessible from a PC, allowing the control and monitoring of the individual laser locks. Its scalability
has been demonstrated by simultaneous stabilization of four lasers on a single cavity. We determined the long-term stability of the system to stay well below \SI{1}{\mega\hertz} over the course of an hour. These parameters make our STCL particularly useful for the laser cooling of many molecular species.

\begin{acknowledgments}
We are indebted to Tilman Pfau for generous support and thank Max M\"ausezahl for fruitful discussions and advice on the amplifiers used to drive the piezo actuators~\cite{Mausezahl2023}. This project has received funding from the European Research Council (ERC) under the European Union’s Horizon 2020 research and innovation programme (Grant agreements No. 949431), Vector Stiftung, the RiSC programme of the Ministry of Science, Research and Arts Baden-W\"urttemberg and Carl Zeiss Foundation.
\end{acknowledgments}

\section*{Data Availability Statement}
The \emph{Python} codes used in this study are openly available in the GitHub repository of the Langen Group~\cite{GithubLangenGroup}.

\appendix

\section{Removing overlapping cavity transmission peaks
from different lasers}\label{appendix:shiftingpeaks}

As shown in the main text, many lasers can be coupled into the same cavity to stabilize them using the STCL. In principle, this only requires sufficient power to reliably detect the transmission peak of each laser. The robustness of the STCL, however, depends on the  size of the region of interest available around each transmission peak to uniquely detect and compensate the frequency fluctuations of the corresponding laser. Therefore the spacing between the individual peaks in the cavity transmission signal effectively limits the maximum number of lockable lasers. In particular, overlapping resonances from different lasers can easily interfere with the successful operation of the frequency stabilization algorithm.

\begin{figure}[bh]
~\\
	\centering
    \includegraphics[scale=.95]{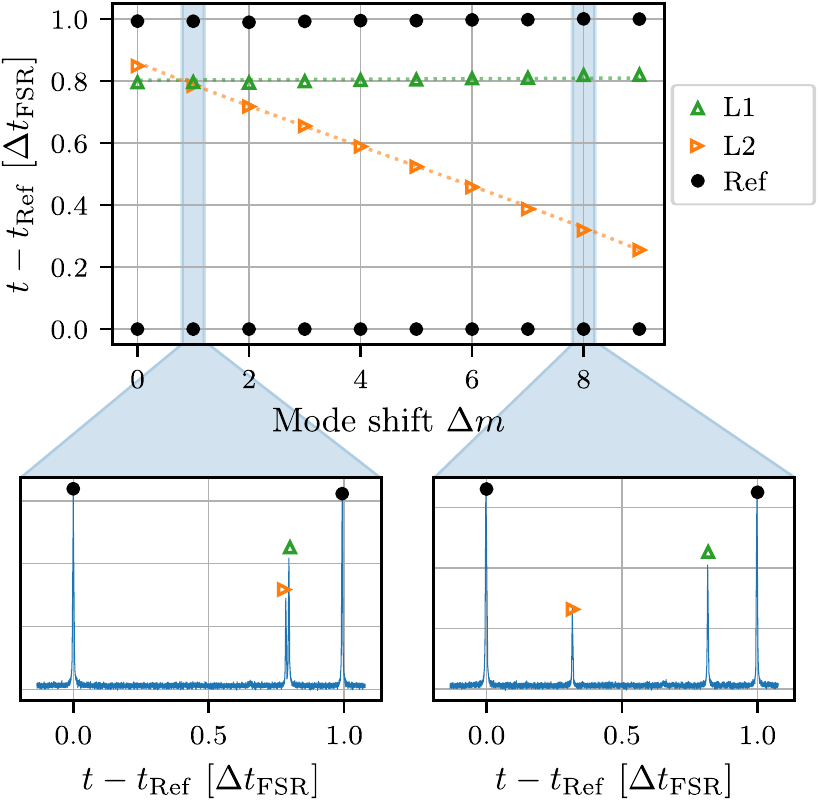}
    \caption{Tunability of relative resonance positions. For the data shown, the piezo of an unstabilized cavity has been shifted over a range of \SI{50}{\volt}. The visible peaks then correspond to resonances with mode numbers that are shifted by $\Delta m$. The relative peak positions for different mode shifts are determined from cavity signals as indicated in the insets. Dotted lines are a guide to the eye. Lasers L1 and L2 have wavelengths of \SI{828}{\nano\meter} and \SI{896}{\nano\meter}, respectively. By offsetting the cavity length, neighboring resonances (left inset, $\Delta m = 1$) of L1 and L2 can be well separated (right inset, $\Delta m = 8$).}
	\label{fig:PiezoOffset}
\end{figure}

For this reason, the cavities used in our work are equipped with an additional piezo actuator to control the overall length of the cavity independent of the scanning piezo actuator that participates in the lock. When scanning the length $L$ of the resonator using this latter piezo actuator, the free spectral range of different lasers depends on their wavelength according to the standing wave condition $L = m \lambda /2, m\in \mathbb{N}$. This wavelength dependence allows us to tune the relative resonance positions of different lasers by choosing an appropriate scan range with the second piezo. This is demonstrated in Fig.~\ref{fig:PiezoOffset}. Effectively, the mode number $m$ is shifted by $\Delta m$ for all lasers. The change in relative position to the reference peak for a laser L can then be modeled as $(\lambda_\text{L}/\lambda_\text{Ref} - 1)\Delta t_\text{FSR}\times \Delta m$. 

As an alternative to this approach, additional transfer cavities can be set up or the frequencies of the lasers can be shifted (e.g. using acousto-optical modulation) before coupling them into the cavity used for the STCL.\\ 

%\nocite{*}
\bibliography{aipsamp,Bibliography}% Produces the bibliography via BibTeX.

\end{document}